\documentclass [a4paper,fleqn, 12pt]{article}
\usepackage{graphicx}
\usepackage[small]{subfigure,epsfig}

\usepackage {amsmath} \usepackage{amssymb} \usepackage{cite}

\begin{document}

\title{Special solutions of high order equation for waves in liquid with gas bubbles}

\author{Nikolay A. Kudryashov \and Dmitry I. Sinelshchikov}

\date{National Research Nuclear University MEPhI (Moscow Engineering Physics Institute), 31 Kashirskoe Shosse, 115409 Moscow, Russian Federation}

\maketitle

\begin{abstract}
A fifth--order nonlinear partial differential equation for the description of nonlinear waves in a liquid with gas bubbles is considered. Special solutions of this equation are studied. Some elliptic and simple periodic traveling waves solution are constructed. Connection of self--similar solutions with Painleve transcendents and their high--order analogous is discussed.
\end{abstract}

\section{Introduction}

Nonlinear ordinary and partial differential equation play an important role in studying various phenomena in modern science (see, e.g. \cite{Ablowitz,Borisov2003,Kudryashov_book,Polyanin,Borisov2013,Borisov2013a}). Particulary, nonlinear partial differential equation are widely used for studying nonlinear waves and structures (see, e.g. \cite{Biswas2006,Polyanin2009,Vitanov2009,Borisov2013,Borisov2013a}). Recently, a fifth order partial differential equation has been derived for description of nonlinear waves in a liquid with gas bubbles \cite{Kudryashov2014}. In the purely dispersive case (i.e. neglecting dissipative effects) this equation has the form:
\begin{equation}
\begin{gathered}
v_{t}+ \alpha v v_{x} +\beta v_{xxx} +\varepsilon \Bigg[\frac{\alpha(2\lambda_{1}+\lambda_{2})}{2}v^{2}v_{x}+(2\beta\alpha-\beta_{1}+3\beta\lambda_{2})v v_{xxx}+\hfill\\+2\beta^{2} v_{xxxxx}+(6\beta\alpha-\beta_{2}+3\beta\lambda_{2}+6\beta\lambda_{1}-2\alpha\lambda_{3})v_{x}v_{xx}
\Bigg]=0. \hfill
\label{eq:extended_KdV_equation}
 \end{gathered}
\end{equation}
where $t$ is the non--dimensional time, $x$ is the non--dimensional Cartesian coordinate, $v$ is the non--dimensional perturbation of gas--liquid mixture density, $\alpha$, $\beta$, $\beta_{1}$ are non--dimensional physical parameters, $\lambda_{1}$, $\lambda_{2}$ and $\lambda_{3}$ are arbitrary parameters introduced by near--identity transformations (see \cite{Kudryashov2014}).  Let us remark that equation \eqref{eq:extended_KdV_equation} is a three--parametric family of equations parameterized by $\lambda_{1}$, $\lambda_{2}$ and $\lambda_{3}$. Note also that equation \eqref{eq:extended_KdV_equation} can be used for description of nonlinear waves on a surface of shallow water \cite{Kudryashov2012} and in other applications \cite{Kudryashov2014}.

There are three integrable nonlinear evolution equations among members of family \eqref{eq:extended_KdV_equation}: the Lax equation (the fifth--order Korteweg--de Vries equation), the Kaup--Kupershmidt equation and the Sawada--Kotera equation. Consequently all of them can be used for the description of nonlinear waves in a liquid with gas bubbles. Although special solutions of the Lax equation have been well studied, less attention has been paid to special solutions of the Kaup--Kupershmidt and Sawada--Kotera equations. Thus there is an interesting problem to study special solutions of these equations.

Below we construct elliptic and simple periodic traveling wave solutions of the Kaup--Kupershmidt and Sawada--Kotera equation. We believe that some of these solutions are obtained for the first time. We also show that self--similar reductions of the Kaup--Kupershmidt and Sawada--Kotera equation are connected with the Painleve transcendents and their high--order analogous \cite{Kudryashov2014b,Kudryashov2014c}.

The rest of this work is organized as follows. In the next section we introduce main equations and transform them to canonical forms. We investigate traveling wave solutions of the Kaup--Kupershmidt and Sawada--Kotera equations in Section \ref{S:tw}.  Section \ref{S:self--similar} is devoted to self--similar solutions of these equations. We also discuss connection of these solutions to the Painleve transcendents and its high--order analogous in Section \ref{S:self--similar}.    In Section \ref{S:conc} we briefly discuss our results.

\section{Main equations\label{S:main_eqs}}

Let us show that equation \eqref{eq:extended_KdV_equation} can be transformed to the Kaup--Kupershmidt and Sawada--Kotera equations.

Indeed, if we assume that
\begin{equation}
\lambda_{1}=\frac{12\alpha\beta-\beta_{1}}{6\beta},\quad \lambda_{2}=\frac{6\alpha\beta+2\beta_{1}}{6\beta}, \quad \lambda_{3}=\frac{34\alpha\beta-4\beta_{2}}{8\alpha},
\end{equation}
then from \eqref{eq:extended_KdV_equation} we get
\begin{equation}
v_{t}+\alpha v v_{x} +\beta v_{xxx} +\varepsilon \Bigg[\frac{5\alpha^{2}}{2}v^{2}v_{x}+5\beta\alpha v v_{xxx}+\frac{25}{2}\alpha\beta v_{x} v_{xx}+2\beta^{2}v_{xxxxx}\Bigg]=0
\label{eq:pre_KK}
\end{equation}
Using in \eqref{eq:pre_KK} scaling transformations
\begin{equation}
t=\frac{1}{2\beta^{2}\varepsilon}t',\quad
x=x'-\frac{1}{20\beta^{2}\varepsilon^{2}}t,\quad
v=\frac{4\beta}{\alpha}v'-\frac{1}{5\alpha\varepsilon}
\end{equation}
we obtain a canonical form of the Kaup--Kupershmidt equation (primes are omitted):
\begin{equation}
v_{t}+20v^{2}v_{x}+10vv_{xxx}+25v_{x}v_{xx}+v_{xxxxx}=0
\label{eq:KK}
\end{equation}

Now let us proceed to the case of the Sawada--Kotera equation. Using the following values of $\lambda_{i}, i=1,2,3$ in \eqref{eq:extended_KdV_equation}:
\begin{equation}
\lambda_{1}=\frac{12\alpha\beta-\beta_{1}}{6\beta},\quad \lambda_{2}=\frac{6\alpha\beta+2\beta_{1}}{6\beta}, \quad \lambda_{3}=\frac{16\alpha\beta-\beta_{2}}{2\alpha},
\end{equation}
we have
\begin{equation}
v_{t}+\alpha v v_{x} +\beta v_{xxx} +\varepsilon \Bigg[\frac{5\alpha^{2}}{2}v^{2}v_{x}+5\beta\alpha v v_{xxx}+5\beta\alpha v_{x} v_{xx}+2\beta^{2}v_{xxxxx}\Bigg]=0
\label{eq:pre_SK}
\end{equation}
Applying scaling transformations
\begin{equation}
t=\frac{1}{2\beta^{2}\varepsilon}t',\quad
x=x'-\frac{1}{20\beta^{2}\varepsilon^{2}}t',\quad
v=\frac{2\beta}{\alpha}v'-\frac{1}{5\alpha\varepsilon}
\end{equation}
from \eqref{eq:pre_SK} we obtain a canonical from of the Sawada--Kotera equation (primes are omitted):
\begin{equation}
v_{t}+5v^{2}v_{x}+5vv_{xxx}+5v_{x}v_{xx}+v_{xxxxx}=0
\label{eq:SK}
\end{equation}

Equations \eqref{eq:KK} and \eqref{eq:SK} were introduced in works \cite{Kaup,Kupershmidt} and \cite{Sawada} correspondingly. Let us remark that solutions of these equations are connected with point vortex and nonlinear special polynomials theories \cite{Kudryashov2007,Kudryashov2011}.
Below we consider traveling wave and self--similar solutions of Eqs.\eqref{eq:KK}, \eqref{eq:SK}.

\section{Traveling wave solutions\label{S:tw}}

In this section we consider traveling wave solutions of Eqs. \eqref{eq:KK}, \eqref{eq:SK}. We construct meromorphic solutions of these equations which are expressed via the Weierstrass elliptic function and trigonometric functions.

Note that elliptic functions are characterized by the number of poles in a parallelogram of periods which is called the order of an elliptic function. Here we construct second--order, fourth--order and six--order traveling wave elliptic solutions of Eqs. \eqref{eq:KK}, \eqref{eq:SK}. We also consider some simple periodic traveling wave solutions of Eqs. \eqref{eq:KK}, \eqref{eq:SK}.

For finding elliptic solutions we use an approach presented in \cite{Kudryashov2010,Kudryashov2011b,Kudryashov2011c,Kudryashov2012,Kudryashov2012a,Kudryashov2014a} This approach allow us to systematically construct elliptic solutions of different orders.

Note that throughout this section we denote by $z_{0}$ an arbitrary constant corresponding to the fact that traveling wave reductions of Eqs. \eqref{eq:KK}, \eqref{eq:SK} are autonomous equations.

Let us consider traveling wave reduction of equation \eqref{eq:KK}. Using the following variables $v(x,t)=-\frac{3}{2}y(z)$, $z=x-C_{0}t$ in \eqref{eq:KK} and integrating once with respect to $z$ we get
\begin{equation}
y_{zzzz}=15 y y_{zz}+\frac{45}{4}y_{z}^{2}-15y^{3}+C_{0}y+C_{1}
\label{eq:KK_tw_reduction}
\end{equation}
where $C_{1}$ is an integration constant.

Equation \eqref{eq:KK_tw_reduction} is F-III equation from the classification by Cosgrove \cite{Cosgrove2000}. The general solution of \eqref{eq:KK_tw_reduction} is expressed via the hyperelliptic function of genus 2. Nevertheless, it is worth constructing some particular solutions of \eqref{eq:KK_tw_reduction} in the explicit form. These solutions can be useful for investigation of waves processes described by \eqref{eq:KK}.

Below, we construct elliptic and simple periodic solutions of \eqref{eq:KK_tw_reduction}. We recover some elliptic solutions obtained in \cite{Kudryashov2014a}. We also obtain some new elliptic and simple periodic solutions of \eqref{eq:KK_tw_reduction}.

Equation \eqref{eq:KK_tw_reduction} admits the following elliptic solutions with one double pole in a parallelogram of periods:
\begin{equation}
\begin{gathered}
y=\wp\{z-z_{0},g_{2},g_{3}\}, \quad g_{2}=\frac{4}{3}C_{0}, \quad g_{3}=-\frac{4}{3}C_{1},\\
y=8\wp\{z-z_{0},g_{2},g_{3}\}, \quad g_{2}=\frac{1}{132}C_{0}, \quad g_{3}=\frac{1}{624}C_{1}
\label{eq:KK_tw_solution_1}
\end{gathered}
\end{equation}

We can construct the following non--trivial fourth--order elliptic solution of \eqref{eq:KK_tw_reduction}:
\begin{equation}
\begin{gathered}
y=\frac{1}{4}\left[\frac{\wp_{z}\{z-z_{0},g_{2},g_{3}\}\pm\frac{1}{12}\sqrt{6C_{1}+12C_{0}A-144A^{3}}}{\wp\{z-z_{0},g_{2},g_{3}\}-A}\right]^{2}-2A,\\
g_{2}=15A^{2}-\frac{C_{0}}{3}, \quad g_{3}=\frac{1}{4}C_{0}A-\frac{1}{24}C_{1}-10A^{3}
\label{eq:KK_tw_solution_3_1}
\end{gathered}
\end{equation}
where $A$ is an arbitrary constant.

\begin{figure}[!t]
\center
\includegraphics[width=6.5cm,height=5cm]{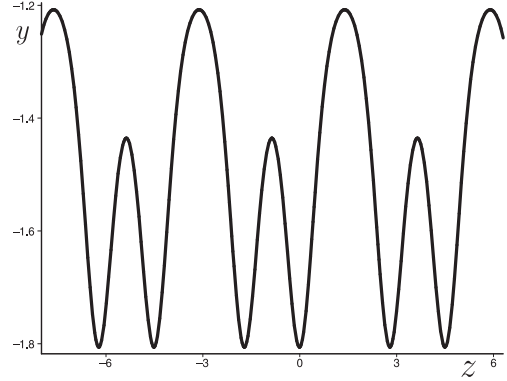}
\caption{Exact solution \eqref{eq:KK_tw_solution_3_1} of  equation \eqref{eq:KK_tw_reduction} at $A=0.903$, $C_{0}=10$ and $C_{1}=0$. }
\label{f1}
\end{figure}

Plot of solution \eqref{eq:KK_tw_solution_3_1} is presented in Fig.\ref{f1}. We can see that solution \eqref{eq:KK_tw_solution_3_1} represents periodic two--crest wave.

The six--order elliptic solution of \eqref{eq:KK_tw_reduction} has the form:
\begin{equation}
\begin{gathered}
y=6\wp\{z-z_{0},g_{2},g_{3}\}+\frac{1}{4}\left[\frac{\wp_{z}\{z-z_{0},g_{2},g_{3}\}+B}{\wp\{z-z_{0},g_{2},g_{3}\}-A}\right]^{2}+\\+
\frac{1}{4}\left[\frac{\wp_{z}\{z-z_{0},g_{2},g_{3}\}-B}{\wp\{z-z_{0},g_{2},g_{3}\}-A}\right]^{2}-4A,\\
g_{2}=12A^{2}-8B_{1}\sqrt{-A}, \quad g_{3}=-B^{2}-8A^{3}-8B(-A)^{3/2}
\label{eq:KK_tw_solution_5}
\end{gathered}
\end{equation}
where $A$ and $B$ are arbitrary constants and the following relations hold
\begin{equation}
C_{0}=1584A^{2}+264B(-A)^{1/2}, \quad C_{1}=1560B^{2}-4992A^{3}-13728B(-A)^{3/2}
\end{equation}

\begin{figure}[!t]
\center
\includegraphics[width=6.5cm,height=5cm]{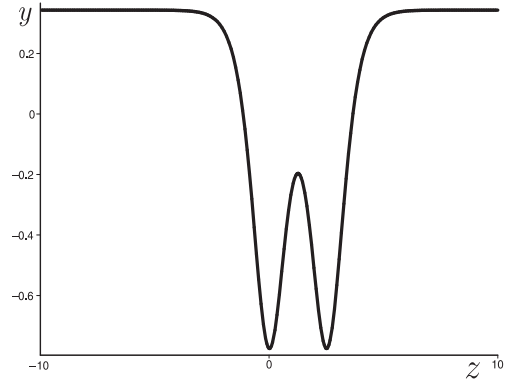}
\caption{Exact solution \eqref{eq:KK_tw_solution_simple_periodic} of  equation \eqref{eq:KK_tw_reduction} at $A=1.01i$ and $T=3i$. }
\label{f1a}
\end{figure}

Now let us consider simple periodic solutions of equation \eqref{eq:KK_tw_reduction}. We know that there are solutions of equation \eqref{eq:KK_tw_reduction} in the form of one--peaked solitary waves. However it is interesting to find solitary wave solutions of equation \eqref{eq:KK_tw_reduction} with more complicated structure. Usually such solutions contain three arbitrary constants. Below, we give an example of such type solution.

We can find the following solitary wave solution of equation \eqref{eq:KK_tw_reduction} containing three arbitrary constants
\begin{equation}
\begin{gathered}
y=\frac{\pi^{2}}{T^{2}}\left[2+\tan^{2}\left\{\frac{\pi(z-z_{0})}{T}\right\}+
\left(\frac{A\tan\left\{\frac{\pi(z-z_{0})}{T}\right\}-1}{\tan\left\{\frac{\pi(z-z_{0})}{T}\right\}+A}\right)^{2}\right]
-\frac{\pi^{2}(3A^{2}+4)}{3T^{2}}
\label{eq:KK_tw_solution_simple_periodic}
\end{gathered}
\end{equation}
where $A$ and $T$ are arbitrary constants. There are also correlations on parameters $C_{0}$ and $C_{1}$:
\begin{equation}
C_{0}=\frac{\pi^{4}}{T^{4}}[16+15A^{2}(3A^{2}+4)], \,
C_{1}=\frac{2\pi^{6}}{T^{6}}\left[-\frac{64}{9}+A^{2}(15A^{4}+30A^{2}+8)\right]
\end{equation}

The plot of solution \eqref{eq:KK_tw_solution_simple_periodic} is presented in Fig.\ref{f1a}.  We can see that solution \eqref{eq:KK_tw_solution_simple_periodic} represents two--crests solitary wave.

We believe that solutions \eqref{eq:KK_tw_solution_5}, \eqref{eq:KK_tw_solution_simple_periodic} are obtained for the first time.

Now let us consider traveling wave reduction of \eqref{eq:SK}. Using the following variables $v(x,t)=-6y(z)$, $z=x-C_{0}t$ in \eqref{eq:SK} and integrating once with respect to $z$ we get
\begin{equation}
y_{zzzz}=30 y y_{zz}-60y^{3}+C_{0}y+C_{1}
\label{eq:SK_tw_reduction}
\end{equation}
where $C_{1}$ is an integration constant. Equation \eqref{eq:SK_tw_reduction} is F-IV equation from the classification by Cosgrove \cite{Cosgrove2000}. The general solution of \eqref{eq:SK_tw_reduction} is expressed via the hyperelliptic function of genus 2. However it is interesting to construct meromorphic solutions of \eqref{eq:SK_tw_reduction} in the explicit form.

Let us find some elliptic solutions of \eqref{eq:SK_tw_reduction} in the explicit form.  We recover some elliptic solutions of \eqref{eq:SK_tw_reduction} presented in \cite{Kudryashov2014a}. We also obtain some new meromorphic solutions of \eqref{eq:SK_tw_reduction}.  Second--order elliptic solutions of \eqref{eq:SK_tw_reduction} have the form
\begin{equation}
\begin{gathered}
y=\wp\{z-z_{0},g_{2},g_{3}\}+h_{0}, \,\, g_{2}=60h_{0}^{2}-\frac{C_{0}}{3}, \,\, g_{3}=80h_{0}^{3}-\frac{C_{0}}{2}h_{0}-\frac{C_{1}}{12},\\
y=2\wp\{z-z_{0},g_{2},g_{3}\},  \quad g_{2}=\frac{C_{0}}{12}, \quad g_{3}=-\frac{C_{1}}{24}
\label{eq:SK_elliptic_second_order}
\end{gathered}
\end{equation}

Equation \eqref{eq:SK_tw_reduction} admits three different fourth--order elliptic solutions. The first one in the following:
\begin{equation}
\begin{gathered}
y=\frac{1}{4}\left[\frac{\wp_{z}\{z-z_{0},g_{2},g_{3}\}+B}{\wp\{z-z_{0},g_{2},g_{3}\}-A}\right]^{2}-A,\quad
g_{2}=\frac{C_{0}}{12},\quad g_{3}=4A^{3}-B^{2}-\frac{C_{0}A}{12}
\label{eq:SK_elliptic_fourht_order_1}
\end{gathered}
\end{equation}
where $A$ and $B$ are arbitrary constants and the following relation holds
\begin{equation}
C_{1}=2C_{0}A+24B^{2}-96A^{3}
\end{equation}

\begin{figure}[!t]
\center
\includegraphics[width=6.5cm,height=5cm]{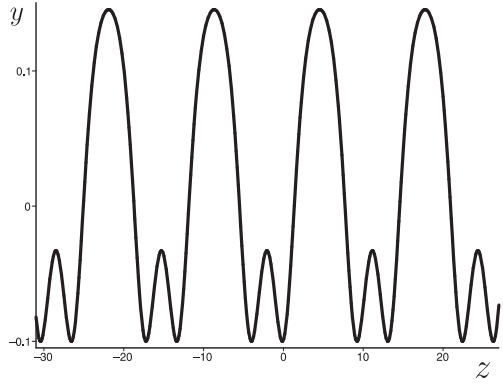}
\caption{Exact solution \eqref{eq:SK_elliptic_fourht_order_1} of  equation \eqref{eq:SK_tw_reduction} at $A=0.1$, $C_{0}=1$ and $B=0.017$}
\label{SK1}
\end{figure}

The second fourth--order elliptic solution has the form
\begin{equation}
\begin{gathered}
y=\frac{1}{4}\left[\frac{\wp_{z}\{z-z_{0},g_{2},g_{3}\}}{\wp\{z-z_{0},g_{2},g_{3}\}-A}\right]^{2}+h_{0},\\
g_{2}=\frac{C_{0}}{12}-15(A+h_{0})(3A+h_{0}),\quad
g_{3}=15Ah_{0}(h_{0}+4A)+49A^{3}-\frac{C_{0}A}{12}
\label{eq:SK_elliptic_fourth_order_3}
\end{gathered}
\end{equation}
where $A$ and $h_{0}$ are arbitrary constants and the following relation holds
\begin{equation}
C_{1}=1440Ah_{0}(3h_{0}+4A)-2C_{0}(2A+3h_{0})+3(768A^{3}+320h_{0}^{3})
\end{equation}
Equation \eqref{eq:SK_tw_reduction} also admits the following fourth--order elliptic solution
\begin{equation}
\begin{gathered}
y=\frac{1}{2}\left[\frac{\wp_{z}\{z-z_{0},g_{2},g_{3}\}}{\wp\{z-z_{0},g_{2},g_{3}\}-A}\right]^{2}-4A,\\
g_{2}=15A^{2}-\frac{C_{0}}{48},\quad
g_{3}=\frac{A}{48}(C_{0}-528A^{2})
\label{eq:SK_elliptic_fourth_order_5}
\end{gathered}
\end{equation}
and the following relation holds
\begin{equation}
C_{1}=4A(192A^{2}-C_{0})
\end{equation}
Solution \eqref{eq:SK_elliptic_fourht_order_1} was obtained in \cite{Kudryashov2014a}. We believe that solutions \eqref{eq:SK_elliptic_fourth_order_3}, \eqref{eq:SK_elliptic_fourth_order_5} are presented for the first time. We present the plot of solution \eqref{eq:SK_elliptic_fourht_order_1} in Fig.\ref{SK1}. We can see that this solution represents two--crest periodic wave.

\begin{figure}[h]
  \centering
 \subfigure[Six--order solution \eqref{eq:SK_elliptic_six_order_3} at $A=1$ and $B=-0.0002$.]{\includegraphics[width=0.49\textwidth]{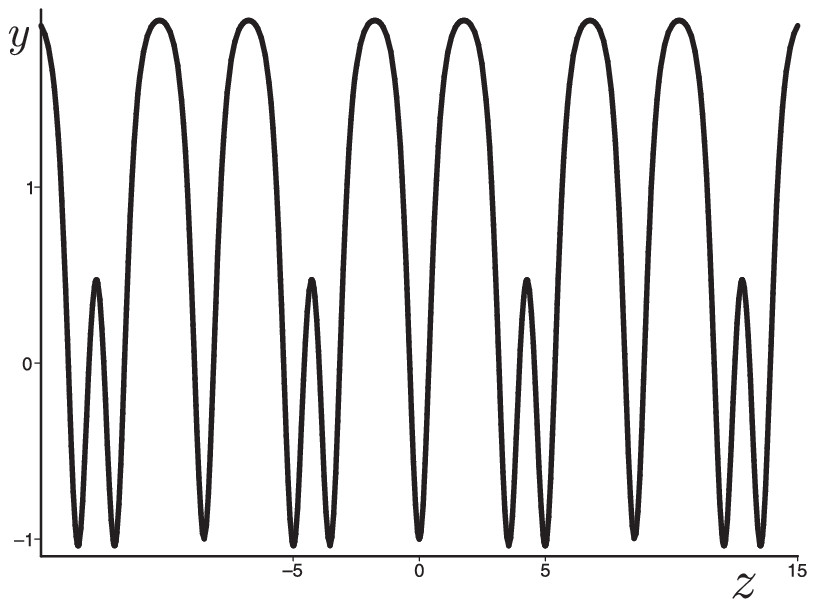}}
 \subfigure[Six--order solution \eqref{eq:SK_elliptic_six_order_5} at $A=1$ and $B=0.0007$.]{\includegraphics[width=0.49\textwidth]{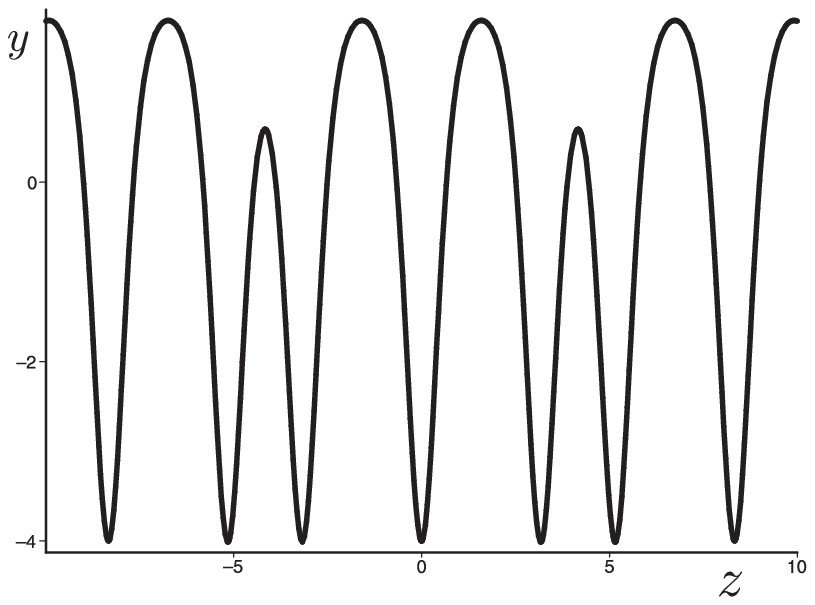}}
    \caption{Six--order elliptic solutions of equation \eqref{eq:SK_tw_reduction}.}
  \label{SK3}
\end{figure}

Let us present six--order elliptic solutions admitted by \eqref{eq:SK_tw_reduction}. There are three different types of non--degenerated six--order elliptic solutions. The first solution has the form
\begin{equation}
\begin{gathered}
y=\frac{1}{4}\left(\left[\frac{\wp_{z}\{z-z_{0},g_{2},g_{3}\}-B}{\wp\{z-z_{0},g_{2},g_{3}\}-A}\right]^{2}
+\left[\frac{\wp_{z}\{z-z_{0},g_{2},g_{3}\}+B}{\wp\{z-z_{0},g_{2},g_{3}\}+A}\right]^{2}\right),\\
g_{2}=\frac{C_{0}}{132},\quad g_{3}=\frac{C_{1}}{312},\quad
A=\frac{\sqrt{33C_{0}}}{132}, \quad B=\frac{\sqrt{-78C_{1}}}{156}
\label{eq:SK_elliptic_six_order_1}
\end{gathered}
\end{equation}
The second six--order elliptic solution is the following
\begin{equation}
\begin{gathered}
y=\frac{1}{4}\left(\left[\frac{\wp_{z}\{z-z_{0},g_{2},g_{3}\}+B}{\wp\{z-z_{0},g_{2},g_{3}\}-A}\right]^{2}
+\left[\frac{\wp_{z}\{z-z_{0},g_{2},g_{3}\}-B}{\wp\{z-z_{0},g_{2},g_{3}\}-A}\right]^{2}\right)-\\-\wp\{z-z_{0},g_{2},g_{3}\}-3A,\\
g_{2}=12A^{2}+4\sqrt{3A}B, \quad g_{3}=-8A^{3}-B^{2}-4\sqrt{3A^{3}}B,\\
C_{0}=144A^{2}+108\sqrt{3A}B, \quad C_{1}=192A^{3}-324B^{2}+72\sqrt{3A^{3}}B
\label{eq:SK_elliptic_six_order_3}
\end{gathered}
\end{equation}
Here $A$ and $B$ are arbitrary constants.

Now we present the third six--order elliptic solution:
\begin{equation}
\begin{gathered}
y=\frac{1}{2}\left(\left[\frac{\wp_{z}\{z-z_{0},g_{2},g_{3}\}+B}{\wp\{z-z_{0},g_{2},g_{3}\}-A}\right]^{2}
+\left[\frac{\wp_{z}\{z-z_{0},g_{2},g_{3}\}-B}{\wp\{z-z_{0},g_{2},g_{3}\}-A}\right]^{2}\right)-\\-2\wp\{z-z_{0},g_{2},g_{3}\}-8A,\\
g_{2}=12A^{2}-4\sqrt{3A}B, \quad g_{3}=-8A^{3}-B^{2}+4\sqrt{3A^{3}}B,\\
C_{0}=144A^{2}+432\sqrt{3A}B, \quad C_{1}=192A^{3}-648B^{2}-1440\sqrt{3A^{3}}B
\label{eq:SK_elliptic_six_order_5}
\end{gathered}
\end{equation}
where $A$ and $B$ are also arbitrary constants.

\begin{figure}[h]
  \centering
 \includegraphics[width=0.49\textwidth]{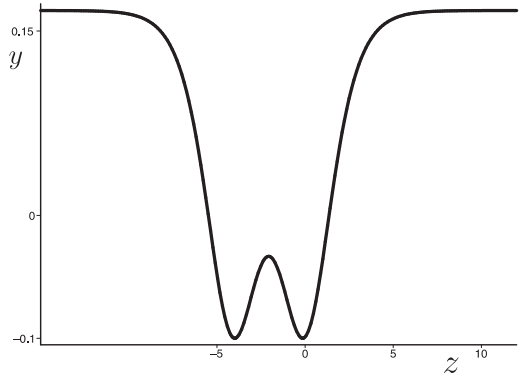}
    \caption{Simple periodic solution \eqref{eq:SK_simple_periodic} of equation \eqref{eq:SK_tw_reduction} at $C_{0}=1$ and $A=0.1$.}
  \label{SK4}
\end{figure}

Note that solution \eqref{eq:SK_elliptic_six_order_1} was obtained in \cite{Kudryashov2014a}. We believe that solutions \eqref{eq:SK_elliptic_six_order_3} and \eqref{eq:SK_elliptic_six_order_5} are new.

Plots of solutions \eqref{eq:SK_elliptic_six_order_3} and \eqref{eq:SK_elliptic_six_order_5} are presented in Fig.\ref{SK3}. We see that these solutions represent three-crest periodic waves.

It is worth considering degenerate cases of above presented fourth--order and six--order elliptic solutions. Note that the Weierstrass elliptic functions degenerates to rational or trigonometric functions in the case $g_{2}^{3}-27\,g_{3}^{2}=0$. We considered all of above presented fourth--order and six--order elliptic solutions assuming that this relation holds. We found that the most interesting case corresponds to solution  \eqref{eq:SK_elliptic_fourht_order_1}.

Indeed, let $B=B_{1}=1/36\sqrt{5184 A^{3}-108C_{0}A+6C_{0}^{3/2}}$ in \eqref{eq:SK_elliptic_fourht_order_1}. Replacing the Weierstrass elliptic function on $k^{2}/3(3\tanh^{2}\{k(z-z_{0})\}-2),\,k=C_{0}^{1/4}/2$ in \eqref{eq:SK_elliptic_fourht_order_1}  we obtain the following simple periodic solution of equation \eqref{eq:SK_tw_reduction}:
\begin{equation}
\begin{gathered}
y=\frac{1}{4}\left[\frac{2k^{3}\tanh\{k(z-z_{0})\}\cosh^{-2}\{k(z-z_{0})\}+B_{1}}
{k^{2}/3(3\tanh^{2}\{k(z-z_{0})\}-2)-A}\right]^{2}-A,\quad k=\frac{C_{0}^{1/4}}{2}
\label{eq:SK_simple_periodic}
\end{gathered}
\end{equation}
Here $A$ is an arbitrary constant. The plot of solutions \eqref{eq:SK_simple_periodic} is presented in Fig.\ref{SK4}. We can see that this solution is a double--picked solitary wave.

We can also construct simple periodic solutions of \eqref{eq:SK_tw_reduction} using an approach from \cite{Kudryashov2010}.  Solution analogous to \eqref{eq:SK_simple_periodic} has the form
\begin{equation}
\begin{gathered}
y=\frac{\pi^{2}}{T^{2}}\left[2+\tan^{2}\left\{\frac{\pi(z-z_{0})}{T}\right\}+
\left(\frac{A\tan\left\{\frac{\pi(z-z_{0})}{T}\right\}-1}{\tan\left\{\frac{\pi(z-z_{0})}{T}\right\}+A}\right)^{2}\right]
-\frac{2\pi^{2}}{3T^{2}}
\label{eq:SK_simple_periodic_1}
\end{gathered}
\end{equation}
where $A$ and $T$ are arbitrary constants and the following relations holds
\begin{equation}
C_{0}=\frac{16\pi^{4}}{T^{4}}, \quad C_{1}=-\frac{64\pi^{6}}{9T^{6}}
\end{equation}

Solution \eqref{eq:SK_simple_periodic_1} represents a double--picked solitary wave as well. To the best of our knowledge solutions \eqref{eq:SK_simple_periodic}, \eqref{eq:SK_simple_periodic_1} are new.

Let us discuss solutions which were obtained above. We see that there are periodic solutions with simple structure i.e. solutions with one crest. Also we have found two--crest and three--crest periodic solutions. Thus we see that there are three type of periodic waves can exist in a liquid with gas bubbles. We believe that possibility of existence two--crest and three--crest periodic waves was discovered for the first time. We have also obtained double--picked solitary wave solutions of equations \eqref{eq:KK_tw_reduction}, \eqref{eq:SK_tw_reduction}. We suppose that possibility of existence of such waves in a liquid with gas bubbles was also shown for the first time.

\section{Self--similar solutions\label{S:self--similar}}
In this section we consider self--similar solutions of equations \eqref{eq:KK} and \eqref{eq:SK}. We discuss relation of these equations to the Painleve equations and their high--order analogous. We also give some special self--similar solutions of these equations.

Let us consider self--similar solutions of the Kaup--Kupershmidt equation. We use the following variables in \eqref{eq:KK}
\begin{equation}
v=-\frac{3}{2}(5t)^{-2/5}w(z),\quad z=x(5t)^{-1/5}
\end{equation}
to obtain
\begin{equation}
w_{zzzzz}-15ww_{zzz}-\frac{75}{2}w_{z}w_{zz}+45w^{2}w_{z}-2w-zw_{z}=0
\label{eq:KK_self_similar}
\end{equation}

Equation \eqref{eq:KK_self_similar} can be presented in the form
\begin{equation}
\left(d_{z}^{3}-3wd_{z}-\frac{3}{2}w_{z}\right)H=0,\quad H[w]=w_{zz}-6w^{2}+\frac{2}{3}z
\label{eq:KK_self_similar_1}
\end{equation}
Here and below we denote by  $d_{z}$ derivative with respect to $z$.

This equation admits the following first integral
\begin{equation}
HH_{zz}-H_{z}^{2}-3wH^{2}=C_{1},
\label{eq:KK_self_similar_fi}
\end{equation}
where $C_{1}$ is an integration constant. Obviously, at $C_{1}=0$ particular solutions of \eqref{eq:KK_self_similar} can be expressed via the first Painleve transcendent
\begin{equation}
H[w]=w_{zz}-6w^{2}+\frac{2}{3}z=0
\label{eq:KK_P1}
\end{equation}
In the case of $C_{1}=1$ one can find a four--parametric family of solutions of equation \eqref{eq:KK_self_similar} which is expressed via the first Painleve transcendent as well \cite{Cosgrove2000}.

Using Miura transformations
\begin{equation}
w=-\frac{1}{3}(2u_{z}-u^{2})
\label{eq:KK_Miura_map}
\end{equation}
from \eqref{eq:KK_self_similar_1} we obtain
\begin{equation}
(d_{z}-u)d_{z}(u_{zzzz}+5u_{z}u_{zz}-5u^{2}u_{zz}-5uu_{z}^{2}+u^{5}-zu-\beta)=0
\end{equation}
Thus we see that solutions of \eqref{eq:KK_self_similar} can be expressed via solutions of the first member of the $K_{2}$ hierarchy \cite{Kudryashov1999}. This equation admits rational solutions which are expressed via nonlinear special polynomials \cite{Kudryashov2007,Kudryashov2011}.

Let us present some rational solutions of \eqref{eq:KK_self_similar}. Using special polynomials from work \cite{Kudryashov2007} and Miura transformation \eqref{eq:KK_Miura_map} we have
\begin{equation}
\begin{gathered}
w=\frac{1}{z^{2}}, \quad w=\frac{8}{z^{2}}, \quad w=\frac{5z^{3}(z^{5}-144)}{(z^{5}+36)^{2}},\\
\end{gathered}
\end{equation}

Let us consider self--similar solutions of the Sawada--Kotera equation.
We use the following variables in \eqref{eq:SK}
\begin{equation}
v=-6(5t)^{-2/5}w(z),\quad z=x(5t)^{-1/5}
\end{equation}
to obtain
\begin{equation}
w_{zzzzz}-30ww_{zzz}-30w_{z}w_{zz}+180w^{2}w_{z}-zw_{z}-2w=0
\label{eq:SK_self_similar}
\end{equation}
Equation \eqref{eq:SK_self_similar} can be presented in the from
\begin{equation}
\left(d_{z}^{3}-24wd_{z}-12w_{z}\right)\left(w_{zz}-3w^{2}+\frac{1}{12}z\right)=0
\label{eq:SK_self_similar_3}
\end{equation}
These equations admits the first integral
\begin{equation}
2HH_{zz}-H_{z}^{2}-24wH^{2}=C_{1}, \quad H[w]=w_{zz}-3w^{2}+\frac{1}{12}z
\label{eq:SK_self_similar_5}
\end{equation}
where $C_{1}$ is an integration constant. Consequently, we see that at $C_{1}=0$ equation \eqref{eq:SK_self_similar} admits solutions which are expressed via the first Painleve transcendent. A four--parameter solution of \eqref{eq:SK_self_similar_5} at $C_{1}=0$ is expressed in terms of the first Painleve transcendent as well \cite{Cosgrove2000}.

The Miura transformation
\begin{equation}
w=\frac{1}{6}(u_{z}+u^{2})
\label{eq:SK_Miura_map}
\end{equation}
connects solutions of \eqref{eq:SK_self_similar_5} with solutions of the first member of the $K_{2}$ hierarchy
\begin{equation}
(d_{z}+2u)d_{z}(u_{zzzz}+5u_{z}u_{zz}-5u^{2}u_{zz}-5uu_{z}^{2}+u^{5}-zu-\beta)=0
\end{equation}
This equation admits rational solutions that are expressed via nonlinear special polynomials \cite{Kudryashov2007,Kudryashov2011}.

We can also construct rational solutions of \eqref{eq:SK_self_similar} using special polynomials from work \cite{Kudryashov2007} and Miura transformation \eqref{eq:SK_Miura_map}. Some of these solutions have the form
\begin{equation}
\begin{gathered}
w=\frac{1}{z^{2}}, \quad w=\frac{2}{z^{2}}, \quad w=\frac{5z^{3}(z^{5}+576)}{(z^{5}-144)^{2}},\quad
w=\frac{7(z^{10}+1152z^{5}+72576)}{z^{2}(z^{5}-504)^{2}}
\end{gathered}
\end{equation}

In this section we have discussed self--similar solutions of \eqref{eq:KK}, \eqref{eq:SK} and their connection to Painleve transcendents and their high--order analogous. We have presented some rational self--similar solutions of \eqref{eq:KK}, \eqref{eq:SK} in the explicit form.

\section{Conclusion\label{S:conc}}
We have considered special solutions of a fifth--order nonlinear evolution equation for description of nonlinear waves in a liquid with gas bubbles. We have shown that this equation is connected with the Kaup--Kupershmidt and Sawada--Kotera equations. We have constructed some new elliptic traveling wave solutions of the Kaup--Kupershmidt and Sawada--Kotera equations. We have also obtained some simple periodic solutions of these equations. We have discussed connection of self--similar solutions of these equations with Painleve transcendents and their high--order analogous. We have shown possibility of existence of complex periodic waves with two and three crests in a liquid with gas bubbles. We have also found possibility of existence of double--peaked solitary waves in a liquid with gas bubbles.

\section*{Acknowledgment}

This research was supported by Russian Science Foundation, project No. 14-11-00258


\begin{thebibliography}{99}

\bibitem{Ablowitz} M.J. Ablowitz, P.A. Clarkson, Solitons, Nonlinear Evolution Equations and Inverse Scattering, Cambridge University Press, Cambridge, 1992.

\bibitem{Borisov2003} A.V. Borisov, I.S. Mamaev, Moder Methods of Integrable Systems Theory, Institute of Computer Sciences, Moscow-Izevsk, 2003.

\bibitem{Kudryashov_book} N.A. Kudryashov, Methods of nonlinear mathematical physics, Publisher hous "Intellekt", Moscow, 2010.

\bibitem{Polyanin} A.D. Polyanin, V.F. Zaitsev, Handbook of Nonlinear Partial Differential Equations, Chapman and Hall/CRC, Boca Raton--London--New York, 2011.

\bibitem{Borisov2013} A. V. Borisov, I.S. Mamaev, Topological analysis of an integrable system related to the rolling of a ball on a sphere, Regul. Chaotic Dyn. 18 (2013) 356--371.

\bibitem{Borisov2013a} A. V. Borisov, I.S. Mamaev, The dynamics of the chaplygin ball with a fluid-filled cavity, Regul. Chaotic Dyn. 18 (2013) 490--496.

\bibitem{Biswas2006} A. Biswas, K. Swapan, Introduction to non-Kerr Law Optical Solitons, Chapman and Hall/CRC, Boca Raton, 2006.


\bibitem{Polyanin2009} A.D. Polyanin, On the nonlinear instability of the solutions of hydrodynamic-type systems, JETP Lett. 90 (2009) 217–221.

\bibitem{Vitanov2009} N.K. Vitanov, I.P. Jordanov, Z.I. Dimitrova, On nonlinear population waves, Appl. Math. Comput. 215 (2009) 2950--2964.

\bibitem{Kudryashov2014} N.A. Kudryashov, D.I. Sinelshchikov, Extended models of non-linear waves in liquid with gas bubbles, Int. J. Non. Linear. Mech. 63 (2014) 31--38.

\bibitem{Kudryashov2012} N.A. Kudryashov, M.B. Soukharev, M.V. Demina, Elliptic traveling waves of the Olver equation, Commun. Nonlinear Sci. Numer. Simul. 17 (2012) 4104--4114.

\bibitem{Kudryashov2014b} N.A. Kudryashov, Higher Painleve transcendents as special solutions of some nonlinear integrable hierarchies, Regul. Chaotic Dyn. 19 (2014) 48--63.

\bibitem{Kudryashov2014c} A.V. Borisov, N.A. Kudryashov, Paul Painleve and his contribution to science, Regul. Chaotic Dyn. 19 (2014) 1-–19.

\bibitem{Kaup} D.J. Kaup, On the inverse scattering problem for cubic eigenvalue problems of the class $\Psi_{xxx}+ 6Q\Psi_{x}+ 6R\Psi= \lambda \Psi$. Stud. Appl. Math. 62 (1980) 189--216.

\bibitem{Kupershmidt} B.A. Kupershmidt, G. Wilson, Modifying Lax equations and the second Hamiltonian structure, Invent. Math. 62 (1980) 403--436.

\bibitem{Sawada} K. Sawada, T. Kotera, A Method for Finding N -Soliton Solutions of the K.d.V. Equation and K.d.V.-Like Equation, Prog. Theor. Phys. 51 (1974) 1355--1367.

\bibitem{Kudryashov2007} Kudryashov, N.A., Transcendents defined by nonlinear fourth-order ordinary differential   equations, {\textit {J. Phys. A. Math. Gen.}}, 1999, vol. 32, no. 6, pp. 999--1013.

\bibitem{Kudryashov2011} M.V. Demina, N.A. Kudryashov, Point vortices and polynomials of the Sawada-Kotera and Kaup-Kupershmidt equations, Regul. Chaotic Dyn. 16 (2011) 562--576.

\bibitem{Kudryashov2010} M.V. Demina, N.A. Kudryashov, From Laurent series to exact meromorphic solutions: The Kawahara equation, Phys. Lett. A. 374 (2010) 4023–-4029.

\bibitem{Kudryashov2011b} M.V. Demina, N.A. Kudryashov, Explicit expressions for meromorphic solutions of autonomous nonlinear ordinary differential equations, Commun. Nonlinear Sci. Numer. Simul. 16 (2011) 1127–1134.

\bibitem{Kudryashov2011c} N.A. Kudryashov, D.I. Sinelshchikov, M.V. Demina, Exact solutions of the generalized Bretherton equation, Phys. Lett. A. 375 (2011) 1074–1079.

\bibitem{Kudryashov2012a} N.A. Kudryashov, D.I. Sinelshchikov, Exact solutions of the Swift-Hohenberg equation with dispersion, Commun. Nonlinear Sci. Numer. Simul. 17 (2012) 26–34.

\bibitem{Kudryashov2014a} M. V Demina, N.A. Kudryashov, Elliptic solutions in the Henon-Heiles model, Commun. Nonlinear Sci. Numer. Simul. 19 (2014) 471--482.

\bibitem{Cosgrove2000} C.M. Cosgrove, Higher-order Painleve Equations in the Polynomial Class I. Bureau Symbol P2, Stud. Appl. Math. 104 (2000) 1--65.

\bibitem{Kudryashov1999} N.A. Kudryashov, Two hierarchies of ordinary differential equations and their properties, Phys. Lett. A. 252 (1999) 173--179.

\end{thebibliography}
\end{document}